# A Google Earth Engine-enabled Python approach to improve identification of anthropogenic palaeo-landscape features

Filippo Brandolini[1*], Guillem Domingo Ribas[1], Andrea Zerboni[2], Sam Turner[1]

[1] Newcastle University - McCord Centre for Landscape, School of History, Classics and Archaeology, Armstrong Building, NE17RU Newcastle upon Tyne (UK), Phone: +44 (0)1912084455
[2] Università degli Studi di Milano - Dipartimento di Scienze della Terra "A. Desio", Via L. Mangiagalli 34, I-20133 Milano (Italy), Phone: +39 (0)250315292

ORCID IDs:

Filippo Brandolini: https://orcid.org/0000-0001-7970-8578
Guillem Domingo Ribas: https://orcid.org/0000-0002-7848-1321
Andrea Zerboni: https://orcid.org/0000-0002-6844-8528
Sam Turner: https://orcid.org/0000-0002-3257-0070

*Corresponding author: filippo.brandolini@newcastle.ac.uk

## Abstract

The necessity of sustainable development for landscapes has emerged as an important theme in recent decades. Current methods take a holistic approach to landscape heritage and promote an interdisciplinary dialogue to facilitate complementary landscape management strategies. With the socio-economic values of the "natural" and "cultural" landscape heritage increasingly recognised worldwide, remote sensing tools are being used more and more to facilitate the recording and management of landscape heritage. Satellite remote sensing technologies have enabled significant improvements in landscape research. The advent of the cloud-based platform of Google Earth Engine has allowed the rapid exploration and processing of satellite imagery such as the Landsat and Copernicus Sentinel datasets. In this paper, the use of Sentinel-2 satellite data in the identification of palaeo-riverscape features has been assessed in the Po Plain, selected because it is characterized by human exploitation since the Mid-Holocene. A multi-temporal approach has been adopted to investigate the potential of satellite imagery to detect buried hydrological and anthropogenic features along with Spectral Index and Spectral Decomposition analysis. This research represents one of the first applications of the GEE Python API in landscape studies. The complete FOSS-cloud protocol proposed here consists of a Python code script developed in Google Colab which could be simply adapted and replicated in different areas of the world.

**Keywords**: Multispectral analysis; Sentinel-2; Spectral decomposition; Python; Riverscape; Fluvial and Alluvial Archaeology; Buried features

# 1.Introduction
## 1.1 Toward a definition of "Landscape Heritage"

Landscapes emerge through complex interrelated natural and cultural processes and consequently encompass rich data pertaining to the long-term interactions between humans and their environments. Over recent millennia, human activities have become progressively more important in shaping geomorphic change [1] to the extent that some scientists argue Earth's history has entered a new epoch, the Anthropocene [2]. In this context, humans are active geomorphological agents, able to modify the physical landscape and shape anthropogenic landscape features [3]. The multi-temporal analysis of landscape dynamics can help identify how human economic development, land use change and population growth have altered natural resources. Past landscape reconstruction enables a better understanding of human resilience to climatic and environmental changes in different periods and locations, and illustrates examples of sustainable development in the past. At the same time, the analysis of historic land use permits the evaluation of human impact on natural environments [4]. The importance of considering landscape's "natural" and "cultural" heritage values together and promoting interdisciplinary approaches to develop conservation strategies has emerged increasingly strongly over the last decade [5,6]. This interdisciplinary perspective is epitomised in the Council of Europe's European Landscape Convention which defines landscape as 'an area, as perceived by people, whose character is the result of the action and interaction of natural and/or human factors' [7]. This international treaty lays out pathways towards sustainable development in landscape based on a balanced and harmonious relationship between social needs, economic activity and the environment.

## 1.2 GIS and Remote Sensing in landscape studies: the satellite 'revolution'

GIS and remote sensing technologies are increasingly recognized as effective tools for the documentation and management of valuable natural and cultural landscape features [8,9]. In particular, satellite remote sensing technologies have provided significant improvements in landscape research and triggered the development of new tools in disciplines including Ecology [10], Geomorphology [11] and Archaeology [12,13]. For instance, in the last 20 years the use of GIS and satellite imagery has dramatically improved the quality of the Historic Landscape Characterisation approach (HLC) [14].
The advent of the cloud-based platform of Google Earth Engine (GEE) has enabled the rapid exploration and processing of more than 40 years of satellite imagery [15]. GEE combines a multi-petabyte catalogue of geospatial datasets and provides a library of algorithms and a powerful Application Programming Interface (API) [16]. GEE eased the access to publicly available satellite imagery and earth observation tools in many branches of scientific research [17–20], opening new opportunities especially for Landscape Heritage applications [21]. Amongst others, GEE users can access Landsat (from 1972) and Sentinel (from 2014) datasets. The highest resolution available in GEE (up to 10 m/pixel) is offered by the Copernicus Sentinel-2 satellite constellation that represents an invaluable free and open data source to support sustainable and cost-effective landscape monitoring [22–24]. Sentinel-2 carries an innovative wide swath high-resolution multispectral imager (MSI) with 13 spectral bands providing extremely useful information for a wide range of applications such as agricultural and forest monitoring [25,26]. Many studies have considered the potential of Sentinel-2 data in the cultural heritage domain at diverse scales of analysis, from single site up to landscape level [27,28], and as a tool for scientific investigation and heritage management and preservation.

GEE can be employed in two main ways: i) via the JavaScript API on the web-based IDE Earth Engine Code Editor, or ii) via Python API on local machines. A third option consists of using the Python API in Google Colaboratory (commonly referred to as "Colab") [29], a Python development environment that runs in the browser using Google Cloud. In origin the Python API did not support any kind of visual output, but this limit has been quickly overcome with the development of new Python modules. Python has proven to be the most compatible and versatile programming language as it supports multi-platform application development. Finally, Python is continuously improved thanks to the implementation of new libraries and modules [30]. Whilst the potential of Python in modelling landscape dynamics has been widely explored [31,32], few publications have so far documented the use of the GEE Python API [33]. In this paper we propose a complete FOSS (Free and Open Source Software) approach to enhance palaeo-landscape feature detection through the GEE Python API in Colab.

## 1.3 Why Riverscapes?

The potential of using the the GEE Python API in Colab has been tested in this paper on riverine landscapes for a number of reasons. Human activities have often relied on river systems, whether for agriculture, navigation or trade purposes. First, fluvial/alluvial environments have been crucial since prehistory owing to the fertility of alluvial landforms and the availability of water supporting settlement, agriculture, mobility and trade [34]. Archaeological investigations have confirmed that over the last 5000 years human activities have profoundly altered the spatial configuration and rate of fluvial and alluvial geomorphic processes, often inducing profound changes to river geomorphology [35]. Riverine landscapes are excellent examples of landscapes which develop through complex relationships between human activities and environmental factors [36–38].

In recent years, remote sensing and satellite imagery have been successfully applied to identify palaeo-geomorphological features (fluvial avulsion, fluvial channels, abandoned meanders, crevasse splays, backswamps) and anthropogenic structures (canals, irrigation systems, artificial levees) in many parts of the world [39–43].

This paper focuses on northern Italy's Po Plain as an ideal test case for the methodology. A huge amount of field- and remotely-sensed geomorphological data are available for the Po Plain and the whole region has been settled and exploited since the Neolithic period. The potential offered by Sentinel-2 imagery has recently been exploited here to map arable land [44]. In this paper we attempt the first Python application of Sentinel-2 data for heritage research in a Mediterranean landscape and illustrate the possibility of detecting and interpreting buried anthropogenic landscape features originating in different phases. The paper demonstrates for the first time that this kind of approach is effective in European fluvial environments.

# 2. Study Area

The Po Plain (Northern Italy) results from the infilling of the depression among the Alps and the Apennines; it is the largest floodplain in Italy. The region forms a natural bridge between the Mediterranean and continental and eastern Europe, and is consequently a key area for understanding environmental and cultural connections between different contexts [45]. People have been closely engaged with fluvial and alluvial dynamics since the region was first colonised and have actively shaped the geomorphology of the basin's rivers since later prehistory.

## 2.1 Geographic and geomorphological background

The Po Plain is situated in a transitional region between the Mediterranean and the European continental climate zones (Fig. 1). As reported in the Köppen classification, the Po Plain is characterised by a range from humid continental (Cfb) to humid subtropical (Cfa) climate [46]. Intense rainfall (700–1200 mm per year) occurs throughout the year and the seasonal pattern of precipitation strongly influences the annual regime of the Po River [47]. The highest rainfall is reached in spring and autumn while the lowest precipitation is usually registered in January and summer (June and July) [48].
The high levels of relative humidity are a consequence of the specific physiography of the plain, surrounded by the Alps and the Apennines, and the influence of the Adriatic Sea [49,50] (Fig. 1). The geomorphological characteristics of the northern and southern sides of the plain differ profoundly [51,52].
The area along the foothills of the Alps is characterized by the presence of Quaternary glacial amphitheatres [53,54] in front of which fluvial fans slowly degrade southwards and eastwards. The fans are interpreted as a result of the mobilisation of glacial and fluvioglacial sediments by rivers which have formed an outwash plain over time [55]. Different phases of alternating depositional and erosional events have resulted in the formation of terraced landforms along the outwash plains. The southern portion of this area consists of a succession of fluvial terraces shaped by the Po River and its tributaries and dating from the Upper Pleistocene to the Holocene [56,57]. Moving eastward, a large portion of the Po Plain and the Friulan-Venetian Plain were built by aggradation processes during the Last Glacial Maximum (LGM, ~22ka - 16ka years BCE)[58,59]. After that phase the Alpine tributaries of the Po River underwent a dramatic phase of incision that caused the formation of terraces and a downstream shift in deposition zones [60]. On the opposing, southern side of the Po Plain, the Apennine watercourses developed an apron of fluvial mega-fans along the boundary of the floodplain. A well-preserved system of Late Pleistocene to Holocene alluvial fans extends northward between the Apennine foothills and the Holocene plain [61,62]. The distal part of alluvial fans present a telescopic shape resulting from alternating aggradation/entrenchment phases tuned by Holocene climatic changes. Each aggradational cycle triggered an incision at the top of the preexisting fan and the progradation of a new fan in a more distal position [63]. Finally, during the Late Holocene, the aggradation of river beds resulted in channel diversions and frequent inundation of flood-prone areas [64]. Additionally, in the eastward portion of the Po Plain and in the Venetian–Friulian area the Late Quaternary floodplain evolved in response to the climate-controlled development of alluvial systems and sea-level changes [65–67].

## 2.2 Environmental History and human settlements

Thanks to its complex settlement and land-management history, the Po Plain represents an ideal laboratory to assess the potentiality of satellite imagery to enhance riverscapes' palaeo-features.
Since the Mid-Holocene (~5–3ka BCE), Neolithic communities settled with increasing intensity in the Po Plain owing to its suitability for agriculture [68]. During the Bronze Age (~1700 – 1150 BCE), the Po Plain witnessed the emergence of proto-urban civilizations – the Terramare culture – that altered the natural fluvial landscape introducing the earliest systems for hydraulic management of the fluvial network and extensive woodland clearance [69–71]. Deforestation and farming development heightened during the Iron Age – the Etruscan period – when agricultural activities became the major land use and farmers the key agents in modifying the landscape [72,73]. Between the 2nd-1st century BCE, the Po Plain was modified deeply following Roman colonisation with the introduction of the centuriation system for agricultural management which entailed the creation of a regular grid of roads, ditches and fields. In this phase, at least 60% of the surface of the Po Plain was deforested and converted into farmland [74,75]. From the 5th century CE, a lack of maintenance of irrigation networks which may have been linked to political disruption associated with the end of the Roman Empire [76], combined with surface instability triggered by a cool climate phase [77], meant that large portions of the Po Plain changed into wetlands [61]. This progressive waterlogging process endured until the beginning of the 10th century CE with significant implications for settlement and farming practices [78]. Between the 10th and 14th centuries CE – corresponding to the Medieval Warm Period [77] – land reclamation intensified owing to an increased demand for arable land alongside general population growth in Europe [79]. At the beginning of the 12th century CE wetland reclamation, the construction of levees and canalisation increased and a series of canals were constructed in the Po Plain for irrigation and navigation [80,81]. In the Renaissance, extensive land and water management activities advanced the process of land reclamation in many coastal and interior wetlands [64,82]. During the Little Ice Age (~1500–1850 CE ca.) deforestation accelerated and reached its peak in the late 1700s, while the construction of embankments was completed during the 19th century CE [77]. Flood defences and drainage systems were further reinforced during the 20th century to reduce the risk of inundation [83]. Human water/land management and natural resource exploitation (e.g. deforestation and quarrying) have been so widespread over the centuries that only a tiny portion of this riverscape can be considered completely 'natural' today [74].

# 3. Material and Methods

The first application of GIS and Remote Sensing techniques to reconstruct the past landscape settings of the Po Plain dates back to the end of the nineties [84]. Nowadays, significant improvements in FOSS software and the increased availability of open-source satellite datasets enable the development of more efficient remote sensing approaches. The mosaic of cultivated fields on the Po Plain changes all the time which makes uniform visual analysis difficult; this heterogeneity also complicates the detection of past riverscape features, as the factors that influence it (crop types, seasonal rainfall, soil moisture) vary in areas with different environmental conditions. For example, variations in the capacity to retain soil moisture are a major factor precluding or enhancing the detection of ancient hydrological features [28,85]. Multi-temporal datasets have the capacity to include diverse land-use/land-cover (LULC) scenarios enabling identification of features that may not be visible on individual images during a particular period of the year [42].

## 3.1 Sentinel-2 dataset

The Sentinel-2 (S2) satellite constellation was developed by the European Space Agency (ESA) in the framework of the European Commission Copernicus Programme [86]. The twin satellites (A and B) of the S2 programme have a 5-days temporal resolution and their multispectral sensors acquire data in 13 separate bands with a spatial resolution up to 10 m (Tab. 1). In this paper we utilize the GEE dataset S2 MSI (MultiSpectral Instrument), Level-1C orthorectified top-of-atmosphere (TOA) reflectance (dataset availability: June 2015 - present).

Buried natural palaeochannels and human structures result in crop marks and soil marks on the surface because they retain a different amount of moisture compared to the surrounding soil. The identification of crop and soils marks from aerial imagery has informed the identification of buried archaeological sites since the 1920s [87–89]. Satellite multispectral images can be more effective in this respect than traditional aerial photography and researchers have identified key bands for the detection of palaeo-landscape features: Visible (0.4 - 0.7 μm), Near Infrared (NIR) (0.7 - 1.4 μm), and Short-Wave Infrared (SWIR) (1.4 - 3 μm) [27,90,91].

Even with the high resolution of modern satellite sensors, the detection of crop marks is affected by several issues, the most important is the phenological stage of the crops [92]. The heterogeneity of the Po Plain farmland and high annual precipitation rates further complicate the recognition of crop marks in the area. Meanwhile soil marks can appear on bare soil as colour changes, easily identifiable after ploughing: differences in soil colour in ploughed farmland highlight traces of past features whether positive (e.g. damper, wetter material from a palaeochannel or former ditch) or negative (e.g. buried natural or artificial levees) [12].

To help overcome this issue, this study adopted a multi-temporal approach by calculating the mean values of bands over two ninety-day periods (January–March and October–December) of each year from 2015–2020. This choice of time-span was driven by two specific factors. The first is related to the increase in intensity and frequency of drought episodes in the Po Plain in the last decade [48]: the longest recorded period of drought lasted from October 2016 to November 2017 [93]. As noted above, changes in soil moisture retention facilitate the detection of buried features such as river palaeochannels or ancient canals especially in severe drought periods. Secondly, autumn and winter are periods of relatively uniform land cover in the Po Plain: ploughing takes place across large areas of arable land, rice paddy fields have not yet been inundated and other winter crops have not yet reached their maximum growth.

The S2 satellite data were accessed through the Python [94] module `geemap` [95] in Colab [29], a serverless Jupyter notebook computational environment for interactive development [96]. The native GEE Python API has relatively limited functionality for visualizing results but the `geemap` Python module was created specifically to fill this gap. Finally, the Python code developed enables the analysis of the S2 filtered image collection through Spectral Index (SI) and Spectral Decomposition (SD) techniques. Each image was exported in Geo.TIFF format in QGIS [97] where the Min/Max values were adjusted with the Cumulative Count Cut tool. Finally, the figures presented in this paper were generated in the QGIS Layout Editor. The Python modules `rasterio` [98] and `matplotlib` [99] were used respectively to create individual plots for each band of the raster and histograms of their values. These representations enabled the analysis to be completed at a higher level of detail in order to identify which output bands yield better performance for the detection of features (Fig. 2).

## 3.2 Spectral indices

SIs for remote sensing purposes consist of mathematical combinations of different bands to enhance particular environmental characteristics. Their use is common in different fields of research, for example in monitoring variations in snow and glacier cover or in disaster prevention and management [100].

In this study, multi-temporal Red-Green-Blue (RGB) colour composites were used to generate two different compositions: RGB (bands 4-3-2), and False Short Wave Infrared Colour (FSWIR, bands 12 - 8 - 4). RGB provides a true-colour visualization, very similar to the human colour perception, while false-colour images enable the identification of areas with different reflectance response to enhance the visibility of anomalies.

Spectral indices that combine NIR and Red channels generally increase the visibility of crop- and soil-marks. Vegetation Indices (VIs) have been widely tested to detect buried structure and fluvial palaeochannels [28,42,91,92,101]. In particular, Agapiou et al. [90] reformulated the NDVI (Normalized Difference Vegetation Index) to elaborate a specific VI for the identification of archaeological remains: the Normalized Archaeological Index (NAI). Focusing on the low-vegetation period of the year, this study adopted spectral indices that could potentially enhance the detection of soil marks including the Bare Soil Index (BSI). The BSI combines Blue (B2), Red (B11), NIR (B8), and SWIR 1 (B4) spectral bands to capture soil variations [102] according to the formula: (($Red+SWIR\ 1$) – ($NIR+Blue$)) / (($Red+SWIR\ 1$) + ($NIR+Blue$)).

The SWIR and the Red bands are employed to quantify the soil mineral composition, while the blue and the near infrared spectral bands enhance the vegetation. In general, the SWIR spectral range is strongly sensitive to soil moisture content enabling the detection of moisture variations in space and time [12]; recent research suggests the SWIR2 band may be valuable for calculating BSI because it seems more sensitive in terms of classification accuracy [103]. For this reason, the SWIR2 band was used in this study to calculate both FSWIR and BSI indices.

## 3.3 Spectral decomposition

Three different spectral decomposition (SD) techniques were used in this study: Hue, Saturation and Value (HSV), Tasselled Cap Transformation (TCT) and Principal Component Analysis (PCA). HSV, TCT and PCA have been successfully employed to detect both archaeological structure and past fluvial features in different environmental contexts [28,42]. Here these three SD approaches were tested to detect past riverscape features in continental environmental conditions.

### 3.3.1 Hue, Saturation and Value (HSV)

HSV (Hue, Saturation, Value, also known as HSB or Hue, Saturation, Brightness) is an alternative representation of the RGB colour space. HSV performs a rotation from the RGB axis and it is characterized by the three relevant properties: 1- nonlinearity, 2- reversibility and 3 - independence of each component from the others [12]. In our Colab Python script code, we calculate HSV through the GEE method `.rgbToHsv()`.

### 3.3.2 Tasselled Cap Transformation (TCT)

The TCT, known also as Kauth-Thomas technique [104], was developed for enhancing spectral information content of satellite data. The TCT consists in a transformation of the original images into a new data set obtained by linear combinations of the original bands. This SD technique is performed on a pixel basis to better represent the underlying structure of the image according to the formula:

$$TC = WTC\,DN + B$$

where *WTc* stands for Weighted Transforming Coefficient (i.e. specific transformation coefficients statistically derived from images and empirical observations), *DN* for Digital Number and *B* for Bias. The transformation *WTc* depends on the sensor considered, because different sensors have different numbers of bands which, in turn, have different spectral responses [12]. There are three composite variables of TCT bands which are routinely adopted: *Brightness* (TCTb, measure of bare soil), *Greenness* (TCTg, measure of vegetation), *Wetness* or *Yellowness* (TCTw, measure of soil and canopy moisture) [105]. To calculate the TCT bands for S2, the WTCs recently defined by Shi and Xu [106] were adopted for their better performance than previous proposed coefficient indexes [107,108]. Finally, in Colab, we computed the TCT components with the `ee.Array` type utilising the Sentinel-2 TCT Coefficients for the 6-Band Image (Blue, Green, Red, NIR, SWIR1, SWIR 2) (Table 2).

### 3.3.3 Principal Component Analysis (PCA)

The PCA transform (also known as the Karhunen-Loeve transform) consists of a linear transformation which decorrelates multivariate data by rotating the axes of the original feature space and outputs uncorrelated data [109]. PCA reduced dimensionality of the data, providing a new series of less correlated bands, limiting the loss of information and enhancing the features of interest [12,28]. In the Python script code the PCA is calculated by diagonalizing the input band correlation matrix through Eigen-analysis (`eigen()`). Only 10-meter resolution bands were employed in PCA.

# 4. Results

To assess the potential of the procedure discussed in this paper, the Python script code was tested at different locations in the Po Plain with well-known archaeological sites. The key points selected to test the script code consist of well-documented areas where anthropogenic activities have altered the pristine alluvial and fluvial geomorphological settings since protohistory. Pre-existing information about the occurrence of buried natural and anthropogenic features provides a unique set of benchmarks to test the method. The case studies (from West to East) are: *Terramara Santa Rosa di Poviglio* (RE), *Valli Nuove di Guastalla* (RE), *Pra' Mantovani* (MN), *Fabbrica dei Soci* (VR), *Santa Maria in Pado Vetere* (FE) and *Altinum* (VE) (Fig. 1).

**Santa Rosa di Poviglio**
The site of Terramara Santa Rosa di Poviglio is a key settlement associated with the Bronze Age Terramare Culture (TC) [110]. The village and its surroundings were delineated through an artificial modification of a pre-existing crevasse splay lobe. The settlement consists of two moated villages delimited by earth ramparts connected to an adjoining river channel through a canal network [111,112]. The earth ramparts are easily visible in all the SI and SD analysis performed as shown in Fig. 3. The two moated villages are particularly evident in the FSWIR and BSI compositions while RGB, HSV and PCA images highlight the presence of a palaeochannel that flows southwards from the TC site. A square-shape feature lies near the southern limit of the Bronze Age village and could be interpreted as a Roman structure related to the centuriation of the surrounding landscape. East of the Santa Rosa site a series of irregular palaeochannels are the result of the early Medieval waterlogging process that affected large portions of the Po Plain. An elliptical structure in the top right corner of the RGB image is likely to be a false positive: it was not detected in the other SI/SD analysis.

**Valli Nuove Guastalla**

This site lies in the Central Po Plain, not far from the Terramara Santa Rosa di Poviglio site, in a portion of the floodplain known as "backswamp". This geomorphological terms refers to the lowest area of floodplains, poorly drained, where finer sediments accumulate after flooding events [35]. As noted above, the period which witnessed the collapse of the Roman Empire was also associated with climatic instability and progressive waterlogging of the Po Plain. Roman farmland of the backswamps was inundated and became a palustrine environment [64]. Valli Nuove Guastalla is a good location to investigate the impact of the processes which occurred between the Roman and the Medieval eras even though the cultivated mosaic of fields precludes clear visibility of soil marks here (Fig. 4). In the RGB image calculated from the S2 seasonal mean values three buried orthogonal axes are visible, remnants of the drainage system created through Roman centuriation. These palaeofeatures are slightly visible also in the FSWIR and PCA images although hardly recognisable in the others. Buried canals and palaeochannels are highlighted in the FSWIR, HSV and PCA images: these features are most likely the results of flood management during Medieval land reclamation activities in the area [64].

**Pra' Mantovani**

The environmental context of the Pra' Mantovani sites is similar to Valli Nuove di Guastalla. Here, recent archaeological surveys [113,114] have registered the presence of Medieval settlements and buried Roman ditches. In all the SI/SD of Fig. 5, an Early Medieval motte is clearly visible almost in the middle of the area. In the surroundings of this archaeological feature, a series of palaeochannels can be recognised. Positive soils marks in the RGB and PCA images highlight irregular rounded features that may be interpreted as buried archaeological structures.

**Fabbrica dei Soci**

This site is one of the most important TC settlements in the Po Plain. In all the SI and SD images the general pattern of the site and the area nearby is clearly detectable (Fig. 6). The Terramara Fabbrica dei Soci presents a regular square-shaped village centred in a complex hydraulic system that distributed the water diverted from a river palaeo-channel in the surrounding fields for irrigation. The water management documented at this site can be considered as paradigmatic for the whole TC [115,116]. Moats, canals and palaeochannels are especially recognisable in the RGB, FSWIR and PCA images. In the HSV image, the shape of the buried palaeochannels is particularly legible, while in the TCT the square-shaped settlement stands out clearly.

**Santa Maria in Pado Vetere**

Santa Maria in Pado Vetere consists of an Early Medieval church located in the area of the former palustrine environment known as *Valli di Comacchio* (FE). These backswamps were completely reclaimed during the 20th century CE [117,118]. The land reclamation works unearthed several archaeological site, in particular the Etruscan harbour of Spina [119,120], Roman villas and the early Medieval church of Santa Maria [121]. The place name "*in Pado Vetere*" derives from the latin "*Padus Vetus*" and indicates the presence of a Po River palaeochannel. This palaeo-riverscape feature is clearly visible in all the SI/SD images (Fig.

7) crossing the area from NW to SE. That course of the Po River flowed close to the Santa Maria church. Buried artificial canals are connected to the *Padus Vetus* and probably they were used for navigation and irrigation purposes. The archaeological area of Spina and the Santa Maria church cemetery are hardly recognisable due to the resolution of the S2 imagery. In all the images, an ample crevasse splay (East sector of the area) was detected; it is likely to be the result of a severe flood event which occurred in post-medieval period. The highly fragmented pattern of the farmland here precludes the visibility of the Po River palaeochannel and all other buried features: a similar situation was observed in the Valli Nuove di Guastalla.

**Altinum**
Altinum was a Roman harbour on the inner margin of the Lagoon of Venice founded in the 1st century BCE. Its inhabitants colonized the northern lagoon islands in the 5th century CE and created the earliest settlement at Venice. This site was particularly suited for testing the Python script code because the features detected could be compared with the results of a study that reconstructed the urban topography and palaeoenvironmental setting of Altinum using near-infrared (NIR) aerial photographs [109]. Traces of buried hydrological features are visible in the area near the Roman city; Altinum was surrounded by a complex network of rivers and canals that can be recognised in all the SI/SD outputs (except in the BSI).

# 5. Discussion

The results of this study show that buried natural and anthropogenic palaeo-riverscape features can be detected using a GEE Python API in Colab.
The period selected to perform the multitemporal analysis (autumn and winter) proved fruitful in terms of detection of soil marks. Buried features (both natural and archaeological) are more visible on bare soil than in cultivated fields, especially in highly mosaicised farmland, as confirmed in the cases of Valli Nuove di Guastalla and Santa Maria Pado Vetere. To identification the best period of visibility it is crucial to take in consideration crop rotation and meteorological conditions. In the Po Plain the choice was particularly strategic because the autumn and winter seasons are characterised by relatively uniform land cover. Moreover, the detection of soil marks is strongly related to the soil moisture retention of buried features. In this regard, the S2 image collection selected includes severe droughts events (e.g. years 2016 and 2017) alternated with higher precipitation rate periods (e.g. year 2018): this alternation of high and low rainfall intensity enabled the calculation of the mean values of multitemporal bands significant for the identification of soil marks.
In all six case studies the best performance with respect to the SI outputs was provided by the RGB combination. Soil marks are particularly evident in the Bronze Age archaeological sites of Terramara Santa Rosa di Poviglio and Terramara Fabbrica dei Soci. FSWIR composition was particularly effective in the identification of palaeochannels and buried canals. On the other hand, the BSI index registered an overall poor performance except in the identification of positive soil marks related to the general shape of the settlement: buried structures such as moats and village perimeter are clearly detectable even in the BSI. In general, the decision to use the SWIR2 in place of SWIR1 in the FSWIR and BSI combinations returned useful results.

Among the SD techniques tested in this study, the HSV outputs enabled clearest identification of palaeochannels; as noted above the HSV consists of an alternative representation of the RGB colour space.

TCT and PCA were suitable for the identification of riverscape palaeo features in RGB combination. TCT was derived by the composition of TCTb, TCTg and TCTw bands and it was absolutely effective in the identification of positive soil marks. The detection of the palaeohydrography was much evident in the PCA obtained by the combination of the 1st, 2nd and 3rd principal components. As expected, the PCA was the most performing method adopted in this research along with the RGB SI composition. In all the case studies, the PCA outputs returned a detailed and clear image of the riverscape palaeo-features, considering that the first two or three principal components encompass nearly 80 to 90% of the original data's variance [122]. Thanks to their capacity of reducing redundant information and highlighting variance for the recognition of individual elements, if we plot the four bands of the PCA separately, some principal components depict a significant contrast between the background and the palaeochannels and buried canals which, in turn, eases substantially the detection of these features [123] (Fig. 9). Just like the RGB combination, whose B3 – Green and B4 – Red bands depict the palaeoenvironmental features with more accuracy, the values of the bands that provide a better performance are those with the values more clustered, as depicted in the histograms (Fig. 10). Nevertheless, the combination of different SI and SD approaches helped in increasing the detection of palaeo-features and decreasing the occurrence of false positives.

Furthermore, it is necessary to keep in mind that, even in outputs with the better performance as RGB and PCA, palaeo-landscape features smaller than 10 m (the maximum S2 band resolution) are hardly recognisable. This limit could soon be overcome with the implementation of higher resolution datasets in the GEE collections.

Both TCT and PCA are commonly considered time consuming methods especially when it is necessary to calculate large amounts of data. The Python script code tested in this research required less than a minute to calculate all the SI and SD outputs for each case study and the process could be run from any devices regardless of the local machine specifications. That is possible because Colab is a hosted Jupyter notebook service that requires no setup to use, while providing free access to computing resources. The synergy between GEE, Python and Colab is extremely effective and versatile: essentially it is only necessary to change the region of interest (ROI) in the code script to calculate the SI/SD outputs in any area of the world. In order to optimise the results is only necessary to adapt the filtered image collection parameters to the peculiar environmental characteristics of the new ROI. Furthermore, the `geemap` Python module enables the interactive visualisation of the outputs directly in Colab: the images could also be stored in Drive storage or downloaded to the local device for further analysis with GIS or graphical software.

# 6. Conclusion

Free and open source datasets of satellite imagery offer considerable opportunities for landscape heritage stakeholders both for recording and monitoring activities. In this paper, a completely FOSS-cloud procedure was developed to enhance the detection of palaeo-landscape features. S2 satellite imagery has been retrieved in the GEE dataset collection and analysed through a Python script code realized in Colab. Furthermore, the same script code enables the SI and SD analysis of the image collection, previously filtered to optimize the visualisation of soil marks in different case studies in the Po Plain. The outputs obtained can be visualized directly in the Colab browser or downloaded via Google Drive for further graphical applications or spatial analysis.

The choice of the autumn-winter period was shown to be effective for the detection of soil marks in the Po Plain. Choosing the right timespan for a multitemporal analysis is crucial and it depends on peculiar environmental characteristics of the ROI.

The highest discrimination capability was observed in RGB and PCA outputs enabling the recording of buried hydrological features. Most of these have been checked through the available geomorphological and archaeological literature; published case studies interpreting the occurrence of buried features served as a benchmark to validate the script code that was developed. Surprisingly, in other cases (e.g. Terramara Santa Rosa di Poviglio), unknown buried structures were detected in this investigation: further terrain surveys will be necessary to confirm the presence of these palaeo-landscape features. In general, the methodology proposed is very effective in the reconstruction of Mid-Late Holocene landscape evolution of the Po Plain. The main advantages of this method consist of: i) being FOSS, all the software used here are open-licensed; ii) working in cloud, no powerful hardware is necessary to run the script code; iii) high adaptability, changing the ROI is possible to calculate SI and SD outputs for any area of the world; iv) very basic coding skills are required to adapt the code to a ROI with different environmental characteristics.

The development of FOSS-cloud procedures such as those described in this paper could support the identification, conservation and management of cultural and natural heritage anywhere around the world. In remote areas or where local heritage is threatened as a result of political instability, climate change or other factors, FOSS-cloud protocols can facilitate the application of new scientific methods and enable the dissemination of and access to scientific information.

# Author Contributions

# Competing Interests

Pre-print paper. This research has been submitted to the *Open Research Europe* open access publishing platform (under review).

# Grant Information

This work was supported by the European Union's Horizon 2020 research and innovation programme under the Marie Skłodowska-Curie Grant agreement ID: 890561 (HiLSS - Historic Landscape and Soil Sustainability). https://cordis.europa.eu/project/id/890561

# Acknowledgements

The authors thank Prof. Qiusheng Wu (The University of Tennessee, Knoxville - USA) for his suggestions during the development of the script code. Part of the research has been defined with the support of the Dipartimento di Scienze della Terra "Ardito Desio" (Università degli Studi di Milano, Italy) in the framework of the project 'Dipartimenti di Eccellenza 2018–2022' (WP4—Risorse del Patrimonio Culturale) - Italian Ministry of Education, University, and Research (MIUR).

# Supplementary Materials

The Python script code has been deposited on Zenodo, DOI: 10.5281/zenodo.4384105

# Figures

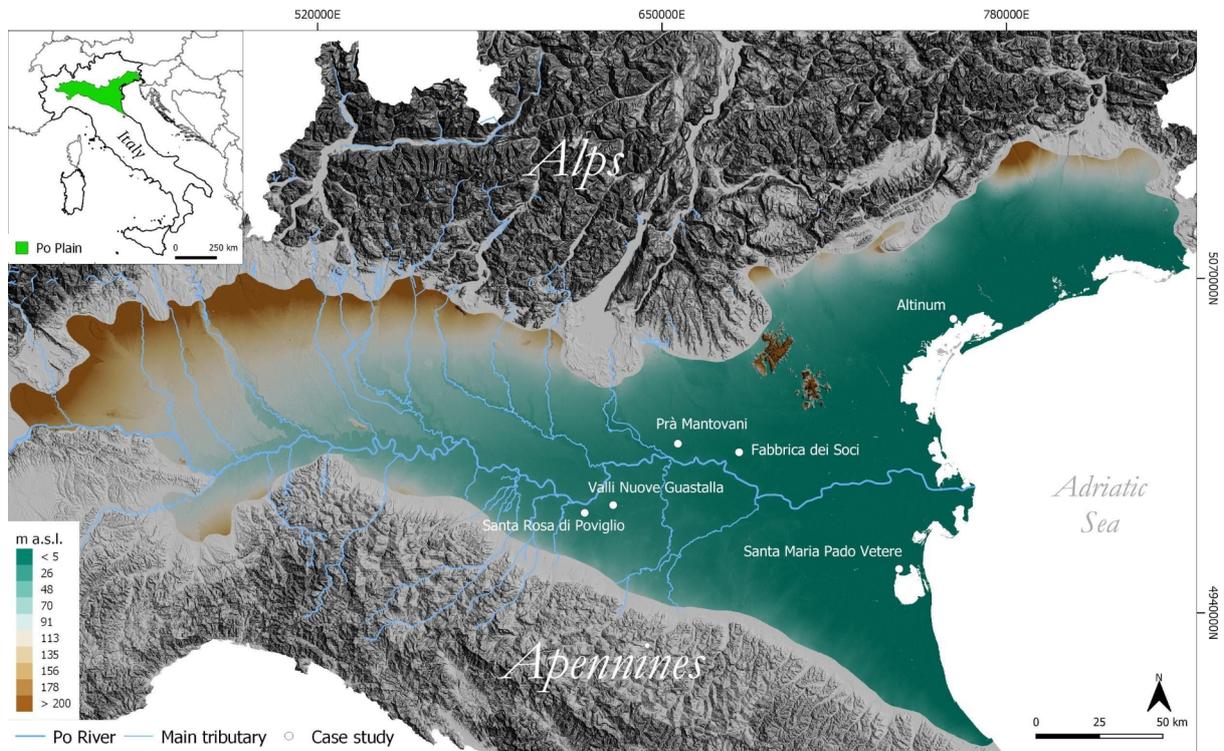

Fig. 1 - Schematic representation of the study area.

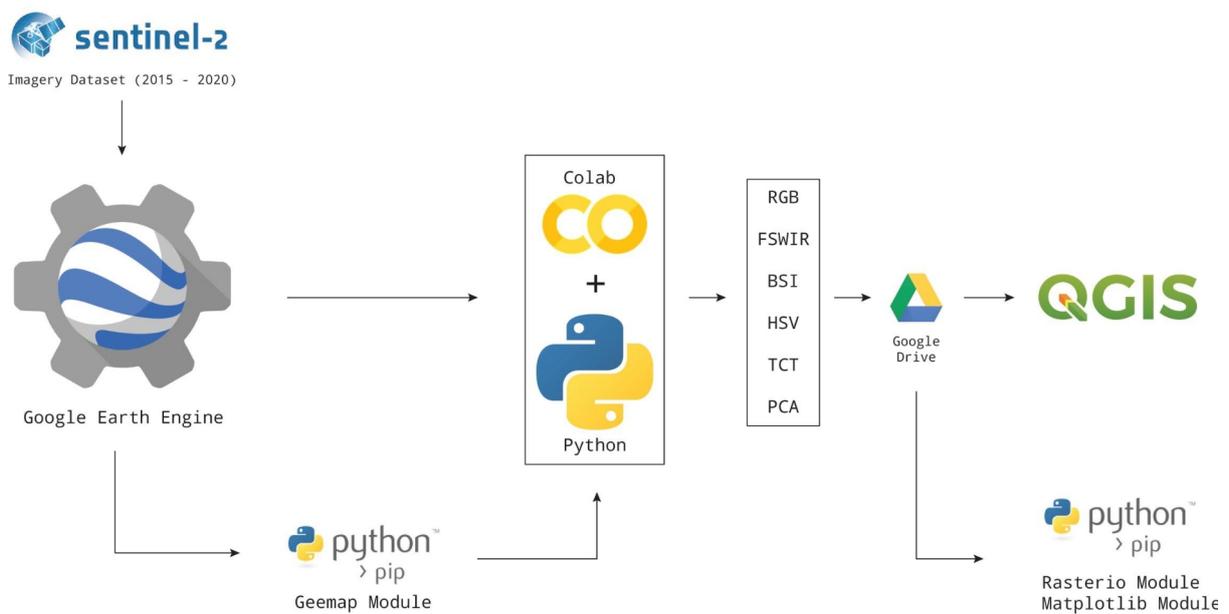

Fig. 2 - FOSS methodological approach adopted in this research.

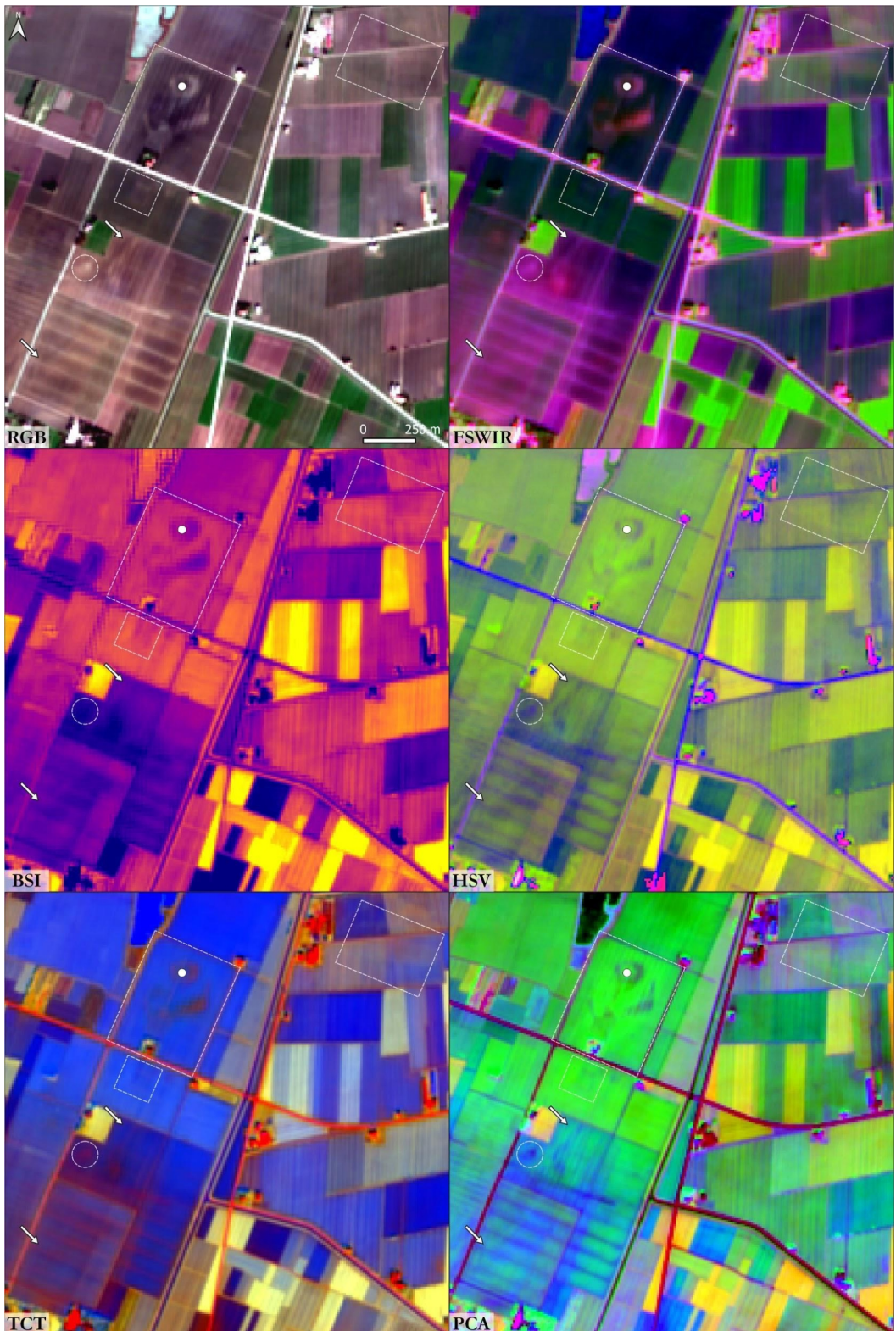

Fig. 3 – Terramara Santa Rosa di Poviglio. The dot is centred on the Bronze Age settlement, the dashed lines frame buried structures and the arrows indicate palaeochannels and canals (Scale and N are indicated in the "RGB" box and are the same for all the other boxes).

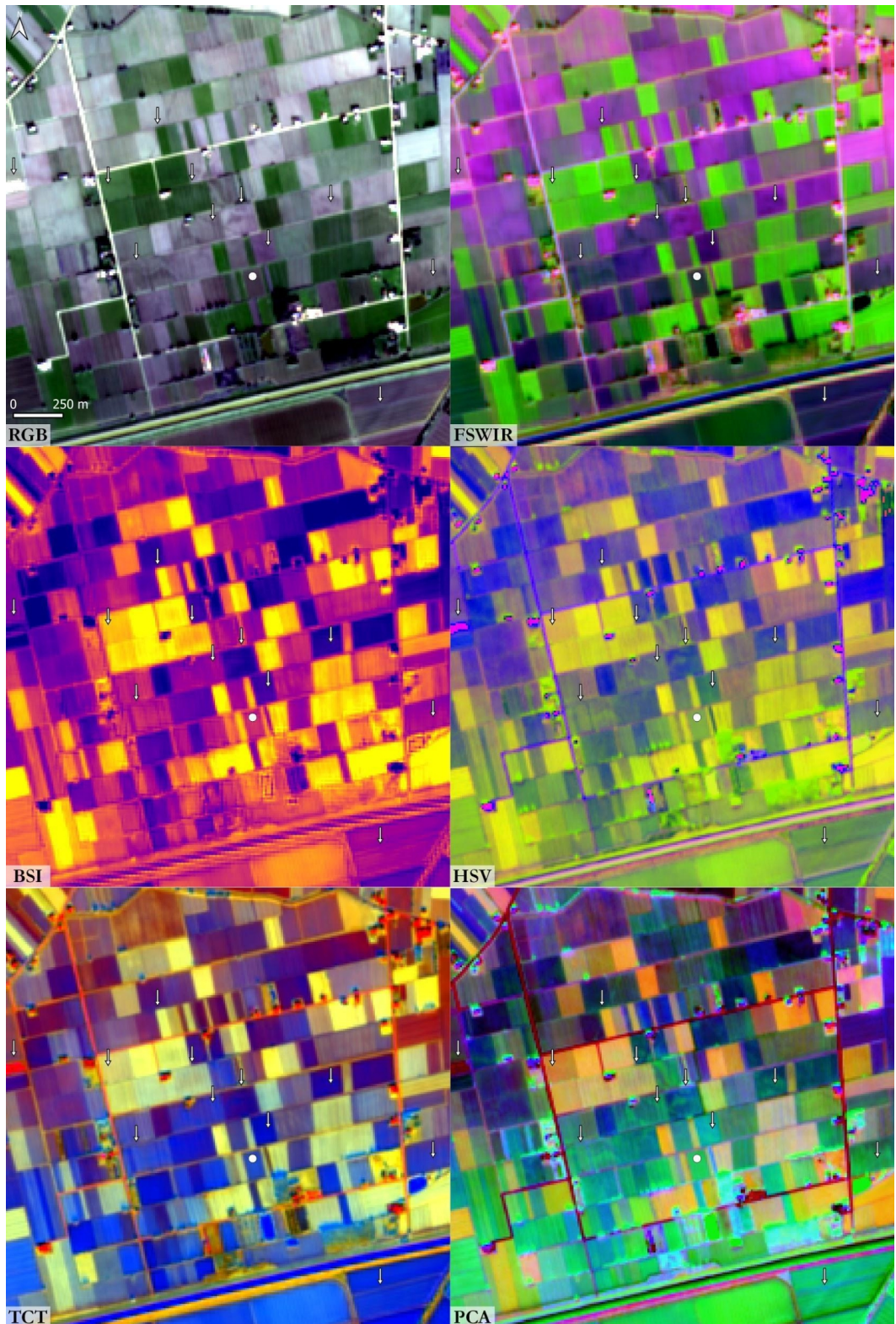

Fig. 4 - Valli Nuove di Guastalla. The dot indicates the case study and the arrows indicate palaeochannels and canals (scale and N are indicated in the "RGB" box and are the same for all the other boxes).

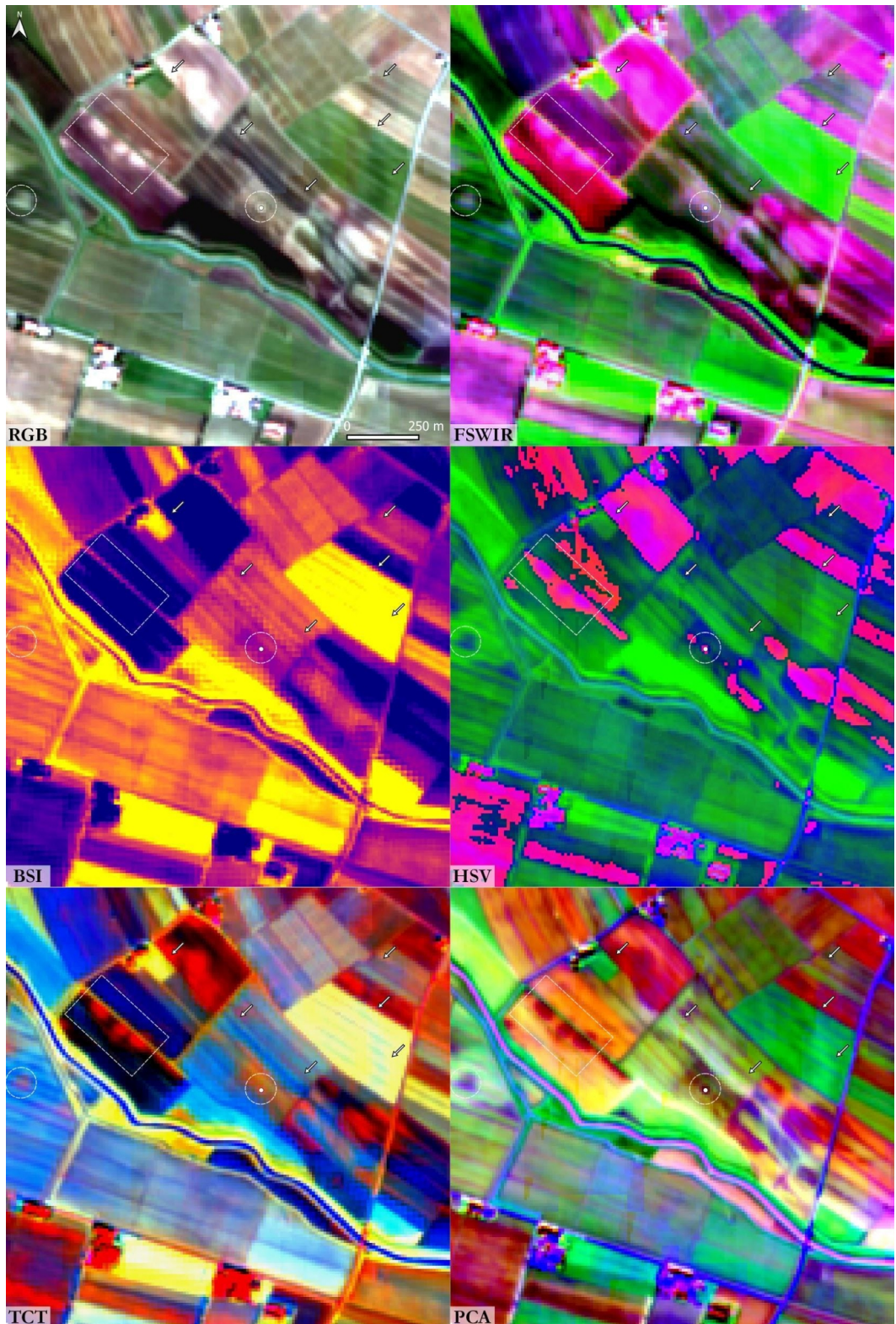

Fig. 5 - Pra' Mantovani. The dot indicates the archaeological site, the dashed lines frame buried structures and the arrows indicate palaeochannels and canals (scale and N are indicated in the "RGB" box and are the same for all the other boxes).

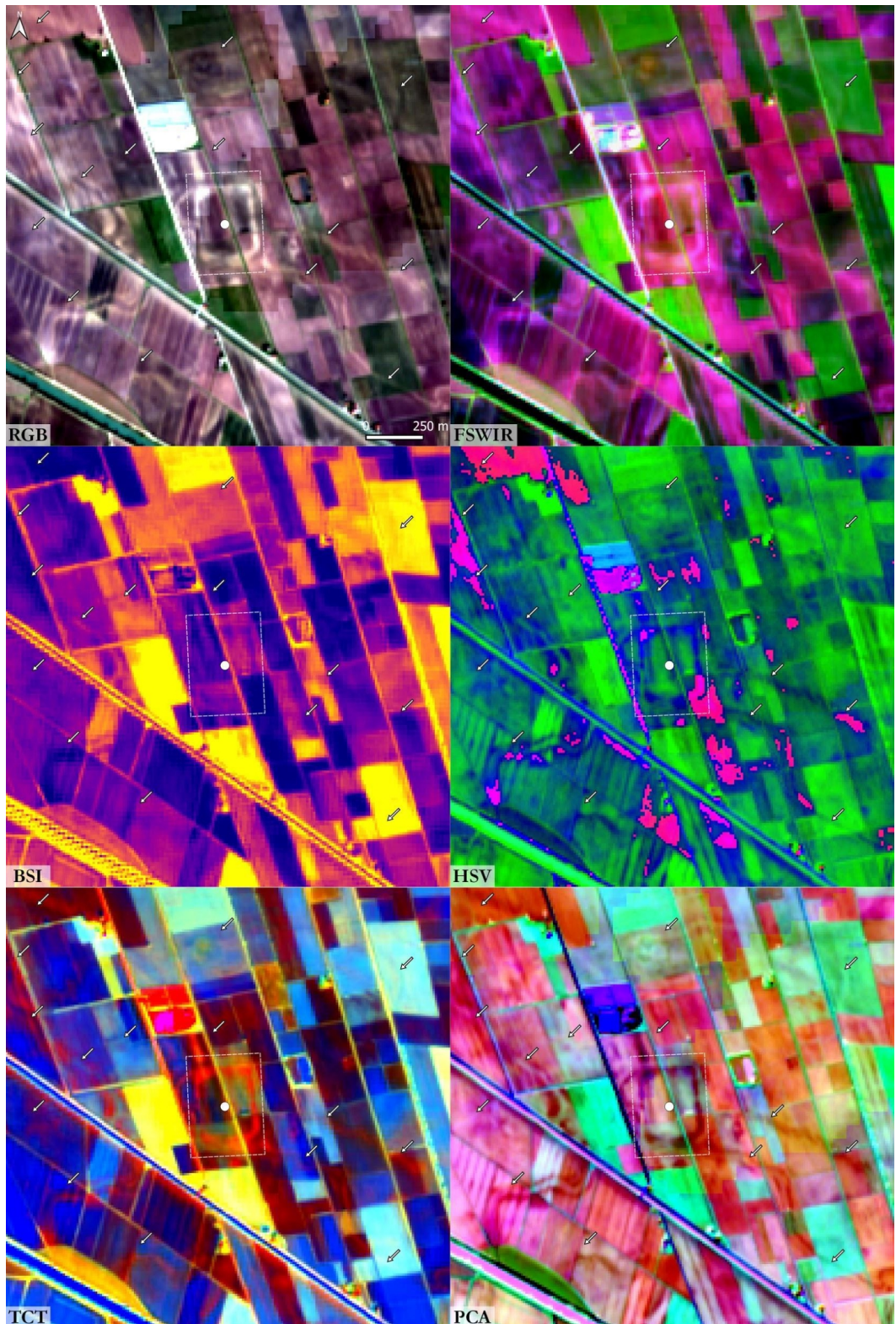

Fig. 6 - Fabbrica dei Soci. The dot indicates the archaeological site, the dashed lines frame buried structures and the arrows indicate palaeochannels and canals (scale and N are indicated in the "RGB" box and are the same for all the other boxes).

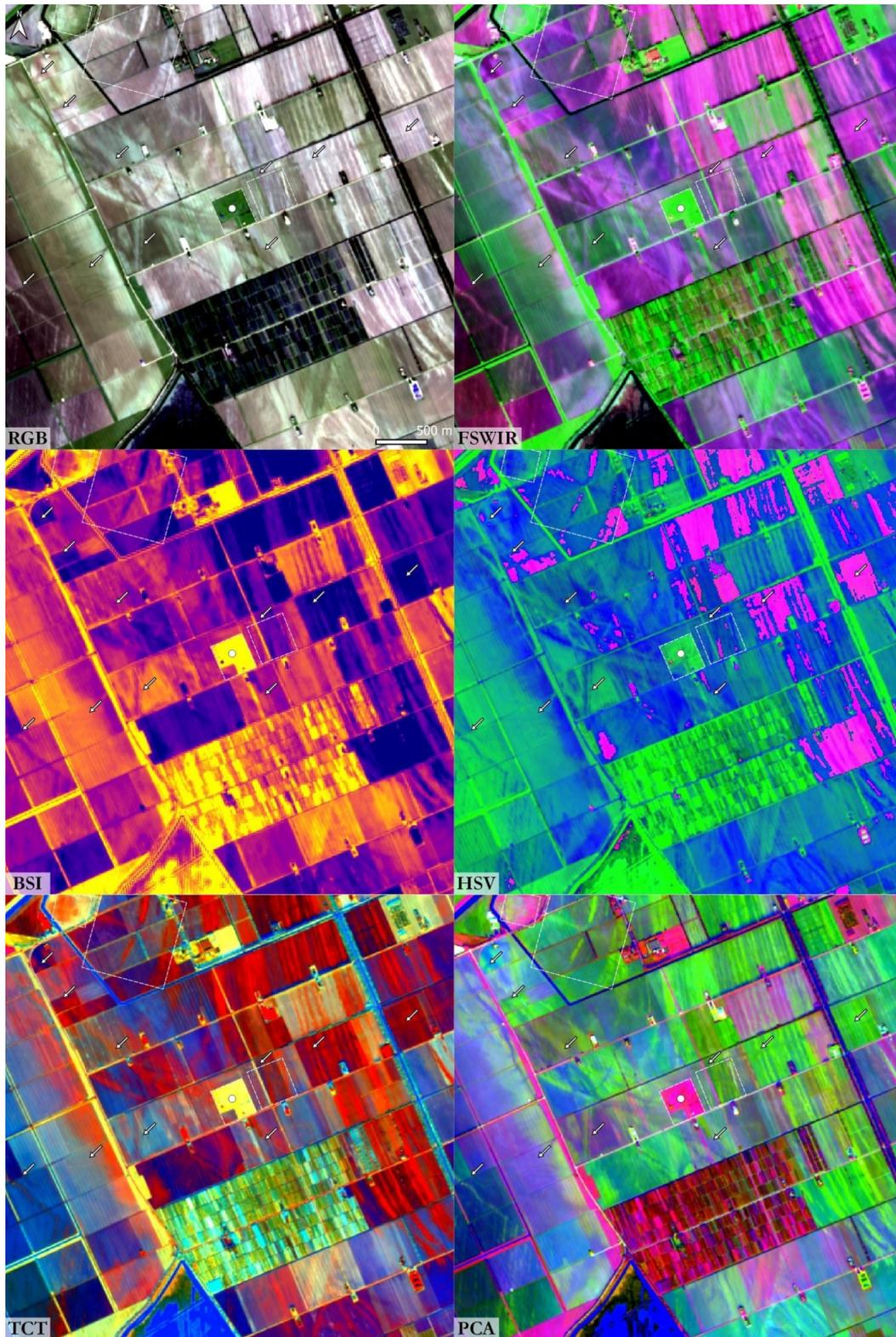

Fig. 7 - Santa Maria in Pado Vetere. The dot indicates the archaeological site, the dashed lines frame buried structures and the arrows indicate palaeochannels and canals (Scale and N are indicated in the "RGB" box and are the same for all the other boxes).

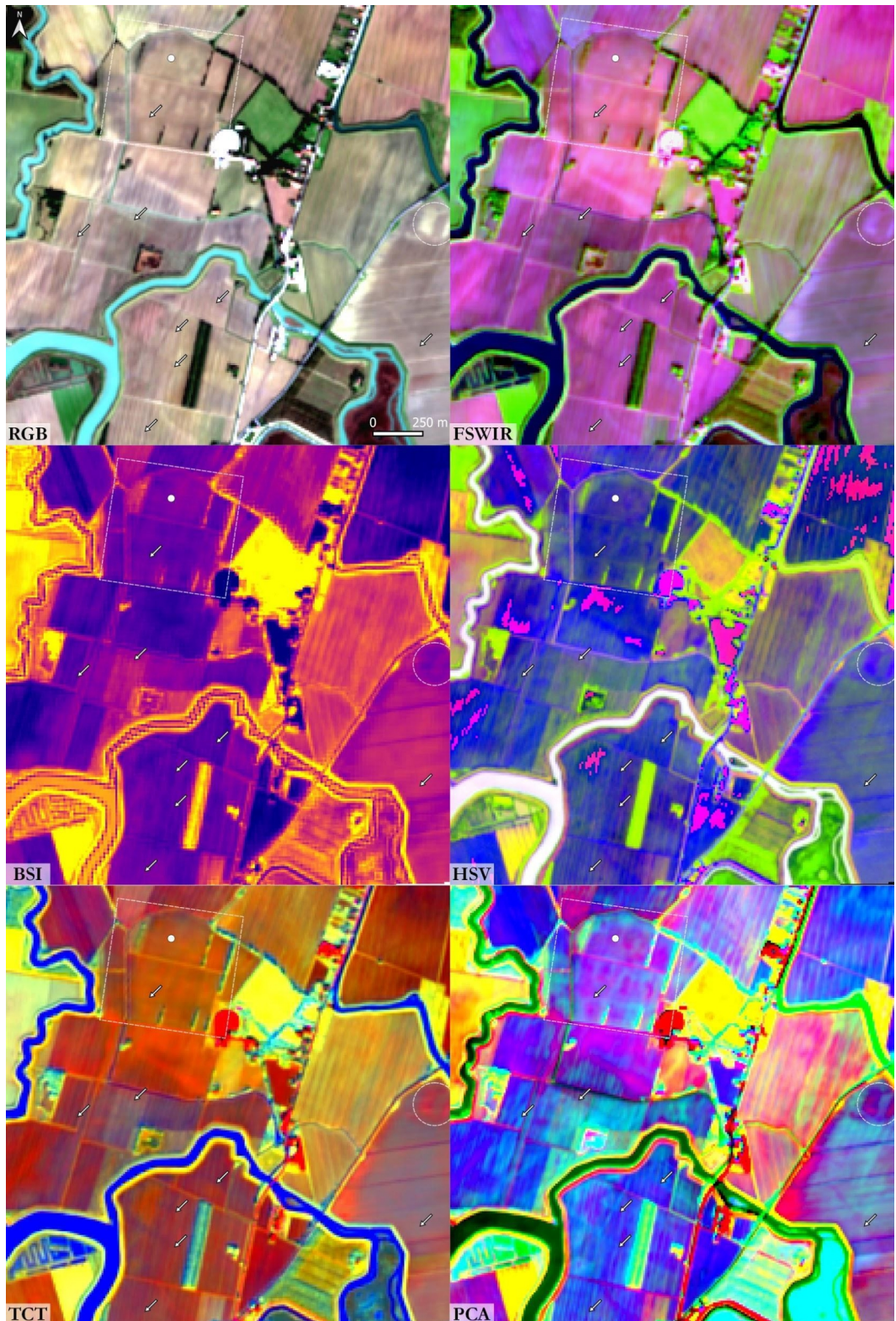

Fig. 8 - Altinum. The dot indicates the archaeological site, the dashed lines frame buried structures and the arrows indicate palaeochannels and canals (Scale and N are indicated in the "RGB" box and are the same for all the other boxes).

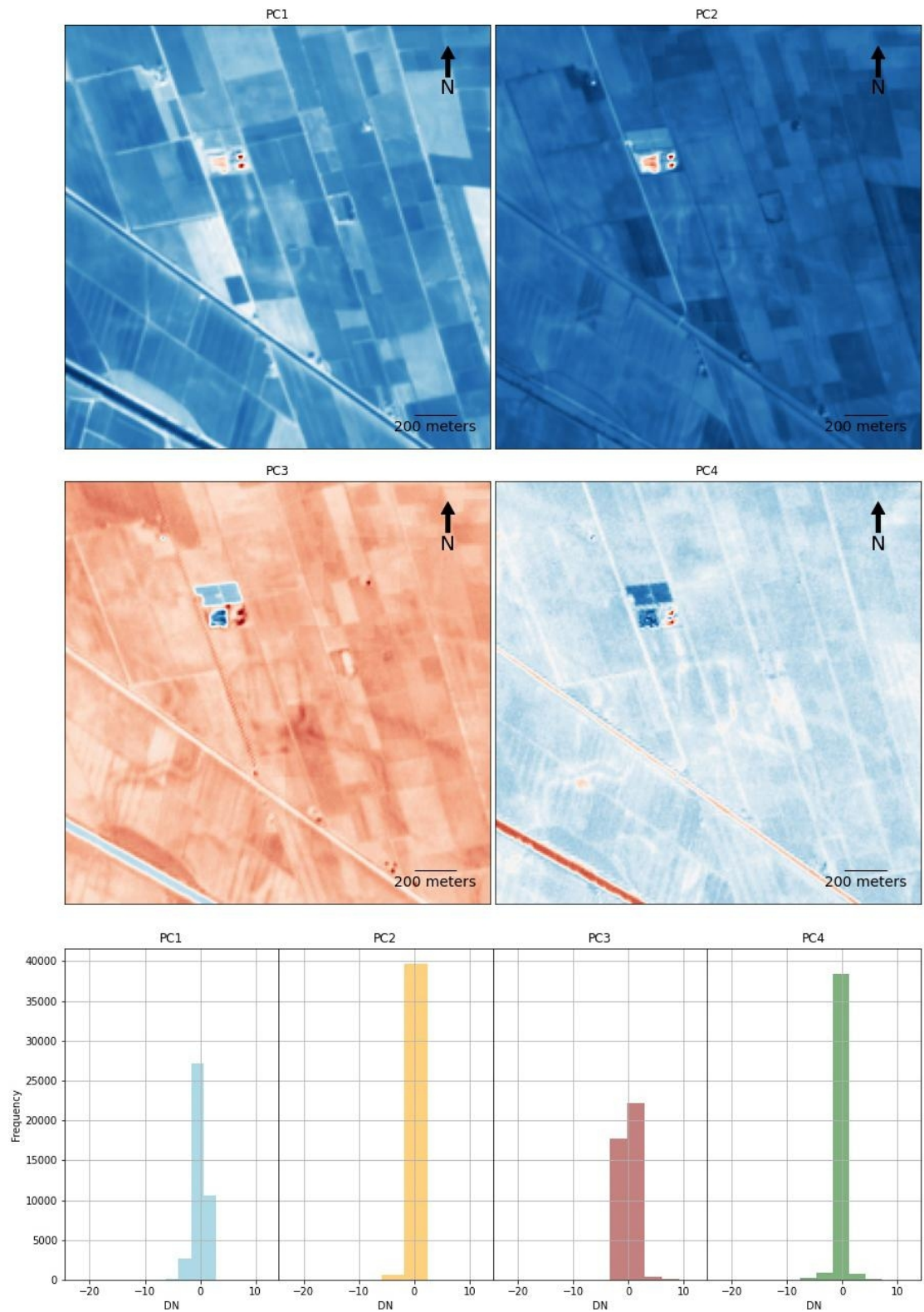

Fig. 9 - Plots (Above) and histograms (below) of the PCA four bands of the *Fabbrica dei Soci* case study (see Fig. 6).

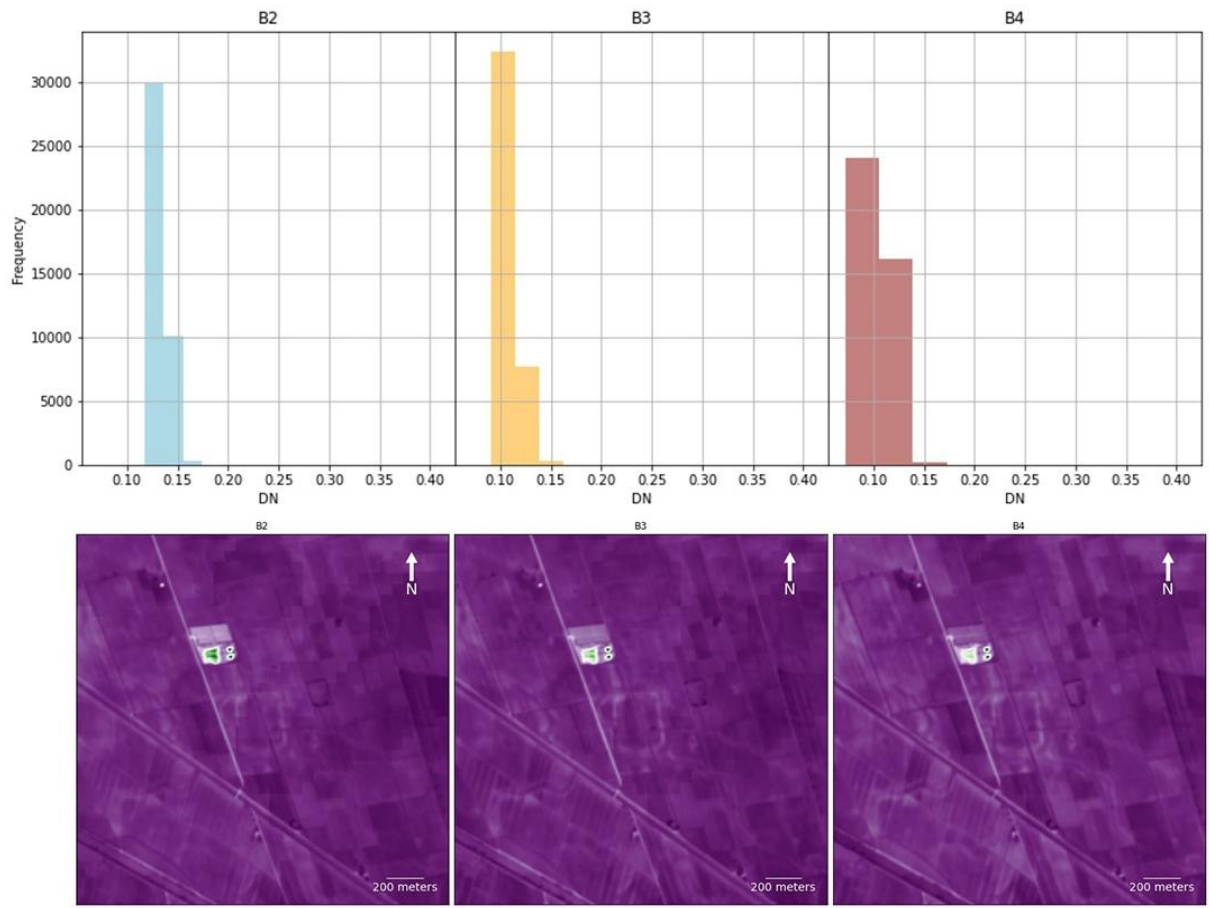

Fig. 10 - Plots (Above) and histograms (below) of the RGB four bands of the *Fabbrica dei Soci* case study (see Fig. 6).

# Tables

| Name | Pixel Size | Wavelength | Description |
|------|------------|------------|-------------|
| B1 | 60 meters | 443.9nm (S2A) / 442.3nm (S2B) | Aerosols |
| B2 | 10 metres | 496.6nm (S2A) / 492.1nm (S2B) | Blue |
| B3 | 10 metres | 560nm (S2A) / 559nm (S2B) | Green |
| B4 | 10 metres | 664.5nm (S2A) / 665nm (S2B) | Red |
| B5 | 20 metres | 703.9nm (S2A) / 703.8nm (S2B) | Red Edge 1 |
| B6 | 20 metres | 740.2nm (S2A) / 739.1nm (S2B) | Red Edge 2 |
| B7 | 20 metres | 782.5nm (S2A) / 779.7nm (S2B) | Red Edge 3 |
| B8 | 10 metres | 835.1nm (S2A) / 833nm (S2B) | NIR |
| B8A | 20 metres | 864.8nm (S2A) / 864nm (S2B) | Red Edge 4 |
| B9 | 60 metres | 945nm (S2A) / 943.2nm (S2B) | Water Vapor |
| B10 | 60 metres | 1373.5nm (S2A) / 1376.9nm (S2B) | Cirrus |
| B11 | 20 metres | 1613.7nm (S2A) / 1610.4nm (S2B) | SWIR 1 |
| B12 | 20 metres | 2202.4nm (S2A) / 2185.7nm (S2B) | SWIR 2 |

Tab. 1 - S2 Satellites bands properties. (https://sentinel.esa.int/web/sentinel/technical-guides/sentinel-2-msi/msi-instrument)

| TCT bands | S2 WTcs |
|-----------|---------|
| TCTb | 0.3510 BLUE + 0.3813 GREEN + 0.3437 RED + 0.7196 NIR + 0.2396 SWIR 1 + 0.1949 SWIR 2 |
| TCTg | -0.3599 BLUE -0.3533 GREEN -0.4734 RED + 0.6633 NIR - 0.0087 SWIR1 -0.2856 SWIR2 |
| TCTw | 0.2578 BLUE + 0.2305 GREEN + 0.0883 RED + 0.1071 NIR -0.7611 SWIR1 -0.5308 SWIR2 |

Tab. 2 - WTcs for S2 defined by [106].